\documentclass[12pt,preprint]{aastex}

\shorttitle{Hard X-ray emission of the microquasar GRO~J1655$-$40.}
\shortauthors{Joinet et al.}

\begin{document}

\title{Hard X-ray emission of the microquasar GRO~J1655$-$40 during the rise of its 2005 outburst.}
\author{A. Joinet\altaffilmark{1}, E. Kalemci\altaffilmark{2}, F. Senziani\altaffilmark{3,4,5}}

\altaffiltext{1}{CESR, 9 Avenue du Colonel Roche, BP4346, 31028 Toulouse, France}
\altaffiltext{2}{Sabanc\i\ University, Orhanl\i\ - Tuzla, \.Istanbul, 34956, Turkey}
\altaffiltext{3}{INAF-IASF Milano, via Bassini 15, 20133 Milano, Italy}
\altaffiltext{4}{Universit\'e Paul Sabatier, 31062 Toulouse, France}
\altaffiltext{5}{Universit\'a degli Studi di Pavia, Dipartimento di Fisica Nucleare e
Teorica, via Bassi 6, 27100 Pavia, Italy}

\begin{abstract}

We present the analysis of the high energy emission of the Galactic black
hole GRO~J1655$-$40 at the beginning of its 2005 outburst.
The data from 458 ks of \textit{INTEGRAL} observations,
 spread over 4 weeks, are analyzed, along with the existing simultaneous
\textit{RXTE} and \textit{Swift} data. The high energy data allow
us to detect the presence of a high energy cut-off and to study its
evolution during the outburst rise. This high energy feature is
generally related to thermal mechanisms in the framework of
Comptonization models from which we can estimate the plasma
parameters. We found an electron temperature of about 30-40 keV and
an optical depth around 1.8-2.1. The high energy cut-off decreased
along with the radio flux, and disappeared as the jet turned off.

\end{abstract}

\section{Introduction}
GRO~J1655$-$40 is a transient Galactic X-ray binary. Since its discovery on
July 27, 1994 with BATSE (Zhang et al. 1994) on-board the
Compton Gamma-ray Observatory (\textit{CGRO}), the source
underwent several outbursts (i.e. : in 1995 see Zhang et al. 1997, in 1996/1997
see Kuulkers et al. 2000 and M\'endez et al. 1998).
It is likely to be a low mass X-ray binary
 and the compact object in this system is probably a black hole
(Bailyn et al. 1995) with a mass estimated at 6.3 M$_ \Sun$ (Greene
et al. 2001)  and
a distance of 3.2 kpc (Hjellming \& Rupen, 1995) from which we based the
calculation of the Eddington Luminosity L$_{edd}$ of the source. A
recent work by Foellmi et al. (2006) places an upper limit of 1.7 kpc
to the distance, but this upper limit is still under debate.
Radio jets were discovered in the mid-1990s and revealed
apparent superluminal motion in opposite directions
(Tingay et al. 1995).

A transient Galactic black hole usually exhibits
a complex spectro-temporal variability with the variation of the accretion
rate. The changes in their  properties  allow to characterise the state
of the source (see McClintock \& Remillard 2006, for a complete description).
In the High Soft State (hereafter HSS), the soft X-ray emission
dominates the spectrum in the form of a blackbody component which is due to the thermal
component from a standard accretion disk.
In the Low Hard State (hereafter LHS), the source is
 characterised by a relatively low flux in the soft X-rays
  ($\la$ 1 keV) and a high flux in the hard X-rays ($\sim$100 keV).
  This is usually interpreted
  (e.g. Shapiro et al. 1976, Narayan \& Yi 1995) as the Comptonization
  of the soft X-ray photons emitted by a cold, geometrically thin disk
  (Shakura \& Sunyaev 1973) by a
  hot plasma surrounding this accretion disk. The hard component is generally
  described by a power-law with a photon index of $\Gamma$=1.4-1.8, and an exponential
cutoff $E_c$ around 100-200 keV. In addition, a Fe K$\alpha$ line at $\sim$ 6.4 keV
and a Compton reflection bump peaking at $\sim$ 30 keV may exist.
 These are signatures of the irradiation of the cold optically thick
 disk by the hard X-rays from the corona (George \& Fabian 1991).
 The LHS is characterized by radio emission which has been shown to be
  consistent with a midly relativistic
 (v $\simeq$ 0.6c) jet (Gallo et al. 2003).
Before entering into the HSS, the source may go through transitional
states also called intermediate states. Homan \& Belloni (2005)
divides the intermediate state into the hard intermediate state
(HIMS) and the soft intermediate state (SIMS) which depend on the
power density spectra and the spectral index. The jet starts to be
quenched in the intermediate state (e.g. Corbel et al. 2004).

We present here an analysis of the spectral evolution of GRO~J1655$-$40
during the rising phase of the 2005 outburst. The analysis is based on
the data taken by the International Gamma-ray Laboratory
({\it INTEGRAL}; Winkler et al. 2003). GRO~J1655$-$40 has been observed by
{\it INTEGRAL} starting from MJD 53425 (\textit{INTEGRAL} revolution 289
on 2005 February 24) to MJD 53448 (\textit{INTEGRAL} revolution 296 on
2005 March 19). The spectral and timing evolution of the source from the
Proportional Counting Array (\textit PCA) on board the Rossi X-ray
 Timing Explorer (\textit{RXTE}) in the 3-30 keV energy range
has been studied in detail in  Shaposhnikov et al. 2007
from MJD 53419 up to MJD 53445. Brocksopp et al. 2006 presented
the data from the Burst Alert Telescope (BAT) on board the
\textit{Swift} observatory in the 14-150 keV energy range
 from MJD 53430 up to MJD 53435. Caballero-Garcia et al. (2007) presented the
 results of the analysis of 4 ToO observations from 27 February to 11
 April of 2005 using JEM-X, ISGRI, and SPI onboard the \textit{INTEGRAL} observatory
 and interpreted it in the framework of several physical models. They come
 to the conclusion that no cut-off is necessary to describe the data in
 the LHS.

We studied the broadband spectral evolution using the
Spectrometer on board the \textit{INTEGRAL} (SPI, Vedrenne et al. 2003)
in the 23-600 keV energy range combined with all the publicly
available data from the PCA and HEXTE detectors on board the \textit{RXTE} observatory,
as well as the BAT detector in order to cover the 3-200 keV energy range.
One of the main interest in studying such a source during the LHS to the
HSS transition is to follow the evolution of the high energy cutoff.
We propose to quantify its value as well as its significance during the
rising phase of the outburst.

We briefly describe the data analysis methods for
\textit{INTEGRAL} (SPI and IBIS/ISGRI), \textit{RXTE} (PCA and HEXTE), and \textit{SWIFT} (BAT)
in Section 2. The scientific results from the spectral modelling
 are presented in Section 3.

\section{Observations and data reduction}

GRO~J1655$-$40 started a new outburst on February 17, 2005 (MJD 53418)
observed by  Markwardt \& Swank (2005) with the
PCA detector onboard the \textit{RXTE} observatory. This reactivation was
confirmed by Torres et al. (2005) who reported the near infrared activity of the
source on Februay 21, 2005. \textit{INTEGRAL} performed follow-up observations
on GRO~J1655$-$40 starting at MJD 53425. Table~\ref{tab:tab1} gives the
details of each \textit{INTEGRAL} revolution and corresponding
quasi simultaneous \textit{RXTE} observations used in this analysis.

\subsection{\textit{INTEGRAL} data reduction}

The data from the SPI detector were reduced as explained in Joinet
et al. (2005) except that only 17 detectors were active
compared to 19 in the beginning of the mission. The SPIROS
V6 algorithm has been used in order to derive the position of
sources detected in the field of view of GRO~J1655-40. We refer to
Joinet et al. 2005 (section 2.1.1) for the method of the spectral
reconstruction. Only pointings for which GRO~J1655$-$40 was at a
distance less than 12 degrees from the central axis were taken into
account for the analysis. We excluded pointings affected by a solar
flare or by exit/entry into the radiation belts. We obtained 458 ks
of useful data during the observation period covered by the
revolutions 289 up to 296 (see Table~\ref{tab:tab1}).
Revolutions 295 \& 296 have been merged and will be named
295+296. The variability time scale of each source was estimated (see
section 3.1) on the basis of both their intensity and their known
temporal behavior. The background flux is stable  within each of the
considered revolution except for revolutions 292 and 293 for which a
variability time scale of one pointing (with a duration of 30-40
minutes) was used. We consider only one normalization parameter per
orbit with an uniformity map determined for the 17 detectors
configuration.

We limited the energy range from 23 to 600 keV for the SPI data and added a
3 \% systematic error to all spectral channels.
The IBIS/ISGRI data corresponding to revolutions 290, 295 \& 296
which were published by Cabarello-Garcia et al. (2007), have also been included
in the analysis. We used the high-level results from them who used the Standard
Off-Line Science Analysis (OSA) 5.1 software package.

\subsection{\textit{RXTE} data reduction}

We also analyzed the public data from the PCA and HEXTE detectors on
board the \textit{RXTE} observatory (Bradt et al. 1993, Rothschild
et al. 1998) data. Table~\ref{tab:tab1} summarizes the set of
\textit{RXTE} observations performed contemporaneously with the
\textit{INTEGRAL} data observation periods except for the revolution
294 for which there are no \textit{INTEGRAL} data available. This
observation period has been divided into three. Several \textit{RXTE}
observations have been merged in order to correspond to one
integrated \textit{INTEGRAL} revolution.

For both instruments the data reduction were performed using the FTOOLS
routines in the HEAsoft software package distributed by NASA's HEASARC
(version 6.0.4).

All available PCUs of the PCA (Bradt et al. 1993) detector have been
used for the data extraction. We added 0.8\% up to 7 keV and 0.4\%
above 7 keV as systematic error (Tomsick et al. 2001)

For HEXTE, we used the response matrix created by the FTOOLS, and applied
the necessary dead time correction (Rothschild et al. 1998). The HEXTE
background is measured throughout the observation by alternating between the
source and background fields every 32 s. The data from the background regions
are merged.

We limited the energy range from 3 to 25 keV, and 16 to 227 keV for the PCA
and HEXTE data, respectively.  \textit{HEXTE} channels were grouped by 2
for channels 16-31, by 4 for channels 32-59, by 10 for channels 60-99
and by 64 for channels 100-227.

\subsection{\textit{Swift} data reduction}

All the publicly available data from the BAT detector on board the
\textit{Swift} observatory  (Gehrels et al. 2004 ,
Barthelmy et al. 2005) covering the rising phase of the 2005
outburst were also analysed (see Table~\ref{tab:tab2}). Some of them were
simultaneous with the {\em INTEGRAL} revolutions (292, 295+296).
 The revolution 292 has been divided into two datasets composed of 7
BAT pointings with a duration of 10.09 ks for 292$_{sw}$-A, and
7.66 ks for 292$_{sw}$-B. We also used the \textit{Swift} data
covering the observation period between the \textit{INTEGRAL} revolutions 292
and 293 ($293_{sw}$) and between revolutions 293 and 295-B
($294_{sw}-A$ and $294_{sw}-B$). We reduced the data using the
standard \textit{Swift} software\footnote{{\em
http://swift.gsfc.nasa.gov/docs/software/lheasoft}} (version 2.4). A
standard filtering was applied in order to discard the data affected
by high background rate and source occultations. For each interval
over which the BAT pointings were unchanged, we extract a background
subtracted spectra together with the response matrix using the
mask-weighting technique. Systematic errors were applied to the
spectra using {\em batphasyserr} BAT FTOOL. For the spectral
analysis, we limited the energy range from 16 to 150 keV.

For both the \textit{RXTE} instruments and the BAT, the normalization factor
was set free with respect to SPI normalization for all fits.

\section{Results}
\subsection{Light curve}

\begin{figure}
\begin{center}
\includegraphics[scale=0.4]{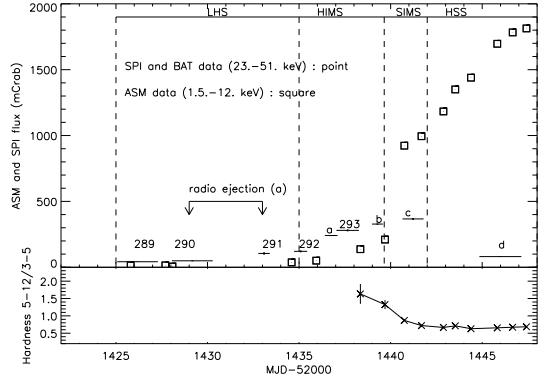}
\end{center}
\caption{Upper panel : ASM (1.5-12 keV), SPI (23-51 keV)
light curves of GRO~J1655$-$40
during the rising phase of the 2005 outburst. The legend of letters
are: a=293$_{sw}$, b=294$_{sw}$-A,c=294$_{sw}$-B, d=295$_{sw}$.
The flux of points a,b,c,d has been extracted from BAT observations. The
different states harbored by the source are summarized on the graph
(see text for the definition): LHS=Low Hard State, HIMS= Hard
Intermediate State, SIMS = Soft Intermediate State, HSS=High Soft State. Two arrows
 indicate (a) the period associated to a radio flare event (see text).
Lower panel : the evolution of the hardness from ASM
in the 3-12 keV energy range which is
defined as the ratio of the ASM fluxes $\Phi$
in two energy ranges $\Phi$ [5.-12. keV]/$\Phi$ [3.-5. keV].}
\label{fig:fig1}
\end{figure}

The flux extraction of GRO~J1655$-$40 from SPI observations was
performed taking into account the hard X-ray sources detected in the
field of view of the source. We used a timescale of one pointing
(whose duration is about 3600 s) for 4U1700-377 and OAO~1657-415 and
two pointings for GX~340+0 and 4U~1705-322. We also considered other
sources (GRO~J1655$-$40, 1E~1740-2942, 4U~1630-47, GX337+00 and
 GX349+2) with a constant flux within each revolution.
As the significance of sources decreases  above 150 keV, we
considered only two sources with a timescale resolution of 3600 s
for 4U1700-77, and one revolution for GRO~J1655$-$40, to extract
fluxes in the 150-600 keV energy range. The light curve of the
source from revolution 289 up to revolution 296 in the 23-51 keV
energy range is shown in Figure~\ref{fig:fig1}. As the source was in
the border of the SPI field of view between \textit{INTEGRAL} revolutions 293
and 295-B, we used the \emph{RXTE} and the
 BAT observations ($294_{sw}$-A, $294_{sw}$-B) to cover this period.
Table~\ref{tab:tab4} gives the SPI and BAT fluxes
in different energy bands.

We also extracted the light curve obtained by the \textit{RXTE}
all-sky monitor ASM in the 1.5-12 keV energy range in the same
period (data taken from the public RXTE
database\footnote{http://xte.mit.edu/lcextrct/asmsel.html}). The
different states harbored by the source were determined by
Shaposhnikov et al. (2007) on the basis of the X-ray properties from
the PCA observations and are summarized in Figure~\ref{fig:fig1}.
The source was in the LHS from revolution 289 up to revolution 292,
it then entered the HIMS from revolution 293 (also observed by the
BAT detector : datasets $293_{sw}$ and $294_{sw}$-A), the SIMS
during the revolution 294-B and was in the HSS from revolution
295-A. Revolution 294 has been divided into two \textit{RXTE} datasets
(294-B and 294-C) in order to follow the evolution of the
high energy cutoff during the SIMS.

As seen from the SPI light curve (Figure~\ref{fig:fig1}), the 23-51
keV flux increased by a factor of 2.9 between MJD 53425 ($\Phi$=42 $\pm$ 2
mCrab) and MJD 53433 ($\Phi$=121 $\pm$ 7 mCrab) while the source was in the
LHS. A radio ejection was observed between MJD 53429-53433
 with a radio peak at 5 GHz (Shaposhnikov et al. 2007).
After this ejection, the X and $\gamma$-ray flux increased
exponentially up to \textit{INTEGRAL} revolution 293 or MJD 53437-53438
($\Phi$=280 $\pm$ 7 mCrab) during which the source was in the HIMS.
As the flux in the 3-25 keV energy range measured from PCA data
increased by a factor 3.9, the flux in the 23-51 keV energy range
decreased by a factor 3.5 between revolution 293 and revolution (295+296)$_{sw}$
indicating the LHS to the HSS transition.

\subsection{Spectral Modeling of the X and $\gamma$-ray data}

The spectra corresponding to each set of data in Table~\ref{tab:tab1} and
\ref{tab:tab2} have been fitted with various
models available in the standard XSPEC 11.3.1 fitting package  (Arnaud 1996).
In all fits, the  iron emission line was modelled by a narrow Gaussian line
fixed at an energy  of 6.4 keV with a free width.
For all models, the inner disk inclination was fixed at $70^\circ$
(Van de Hooft et al. 1998). We also consider the emission from a multicolor
 disk blackbody (DISKBB in XSPEC, Mitsuda et al. 1984). We account for the
interstellar absorption (PHABS in XSPEC) using a hydrogen
column density $N_H = 7 \times 10^{21} \rm \, cm^{-2}$ for most of the observations (see
\S~3.2.1 for exceptions) which is based on the values constrained
by XMM-Newton observation during the 2005 outburst (Diaz Trigo et
al. 2007).

\subsubsection{Power law with a cutoff}

First, a power-law component is added to the base model described above.
The best fit parameters are presented in Table~\ref{tab:tab3}. During the LHS
(revolutions 289, 290, 291, 292),  we found a constant spectral index
 of $\Gamma$ $\simeq$ 1.47 (with a reduced $\chi^2$ ranging from 1.7 up to 2.0).

 During revolutions 289, 290 and 291, the fits are
significantly improved, as seen from the reduced $\chi ^2$ values
(with an F-test probability less than $10^{-9}$),
 by adding a high energy cutoff component (see Table~\ref{tab:tab3}).
A constant spectral index $\Gamma$ of about 1.33-1.36 has been
obtained, which is almost the same value found by
Shaposhnikov et al. (2007) ($1.35 \pm 0.03$). The high energy cutoff
of $E_c = 231^{+94}_{-50}$ keV is consistent with the range value of
163-214 keV presented in Shaposhnikov et al. (2007) for revolution
289. The SPI data that we analyzed for all the outburst allow to
constrain its value and to precisely describe its evolution which
has not been done in Shaposhnikov et al. (2007). However we
reach a conclusion contrary to Caballero-Garcia et al. (2007) which
show that no cutoff was required to describe the data corresponding
to revolution 290. We fit the data set corresponding to
revolution 290 using
  PCA, ISGRI and SPI data with the following
  model : PHABS(DISKBB+POWERLAW)HIGHECUT in order to compare
  our results with the ones derived by Caballero-Garcia et al. (2007).
  We found a high
  energy cut-off powerlaw constrained to a value
  of 22$^{+6}_{-8}$ keV and a folding energy of
  $E_{fold}$=193$^{+14}_{-21}$ keV with a reduced $\chi ^2$ of 0.83 (87 degrees of freedom or
  dof).
  By fitting ISGRI, HEXTE and PCA data, we found a folding energy
  of $E_{fold}$=253$^{+34}_{-31}$ keV and a photon index of
  $\Gamma$ =1.40$^{+0.02}_{-0.02}$  (with a reduced $\chi ^2$ of 1.43 (90 dof)).

For revolution 292, the high energy cutoff is constrained to a value of
$E_c$=$187^{+22}_{-20}$ keV. BAT data were combined with the
simultaneous PCA, HEXTE and SPI data corresponding to the
\textit{INTEGRAL} revolution 292 in order to check the cross
calibration between all instruments.
During revolutions 289-292, the disc component was very weak.
We then notice a decrease of the high energy cutoff down to $E_c$=$87^{+4}_{-5}$ keV during
revolution 294a as the source changes into the HIMS (Shaposhnikov et al.
2007). The disc component also got stronger, while the power law
index remained constant.

By fitting simultaneously BAT, PCA and HEXTE data for revolution 294
  (datasets 294-B and 294$_{sw}$-B), when the source has entered the
  SIMS, a pure powerlaw results in a reduced $\chi ^2$ of
  1.16 (118 dof). Adding a high-energy cutoff results in significant
  reduction in reduced $\chi ^2$ (0.99, 117 dof) but the folding
  energy is not well constrained ($E_{fold}$=$302^{+225}_{-81}$ keV), and strongly depends on
  the power-law index. The disk component also starts to dominate the energy
  spectrum as shown from the clear changes in the $\phi _p / \phi _b$ ratio
(Table~\ref{tab:tab3}) (where $\phi _p $ is the powerlaw flux and
$\phi _b$ the blackbody flux in the  2-20 keV energy range). This
ratio decreases by a factor 13 between revolution 293 and 294-B
due to a large increase of the disk component $\phi
_b$ by a factor of 21. The contribution of the hard component with respect to the disk
component is lower than 50\% during revolution 294-B. Moreover, the
photon index becomes steeper with $\Gamma \simeq 2.1$, indicating a
spectral transition.

\begin{figure}
\begin{center}
\includegraphics[scale=0.4]{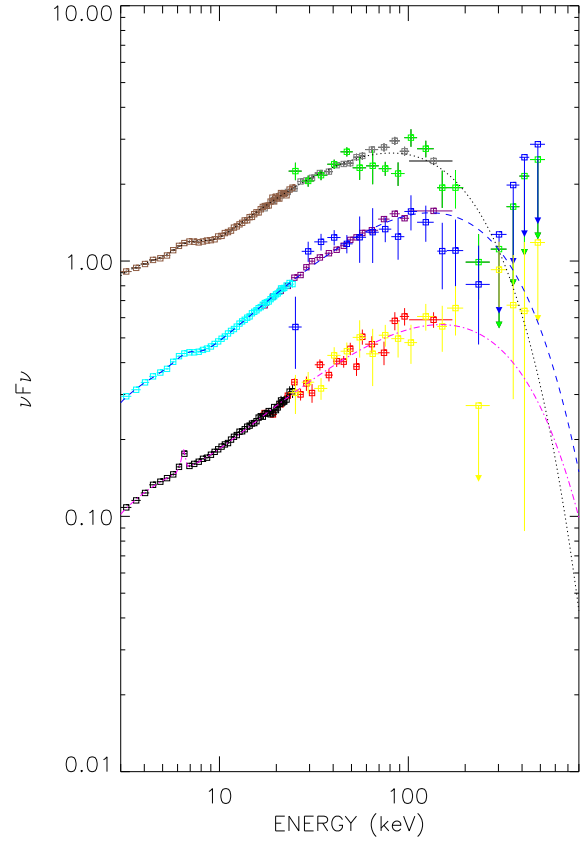}
\end{center}
\caption{Spectra of GRO~J1655$-$40 from
PCA, SPI and HEXTE data fitted simultaneously
with the model described in Table 4.
PCA data for revolutions 289 (black), 292(indigo), 293 (brown);
HEXTE data for revolutions 289 (red), 292(violet), 293 (grey);
SPI data for revolutions 289 (yellow), 292(blue), 293 (green).}
\label{fig:fig4}
\end{figure}
\begin{figure}
\begin{center}
\includegraphics[scale=0.4]{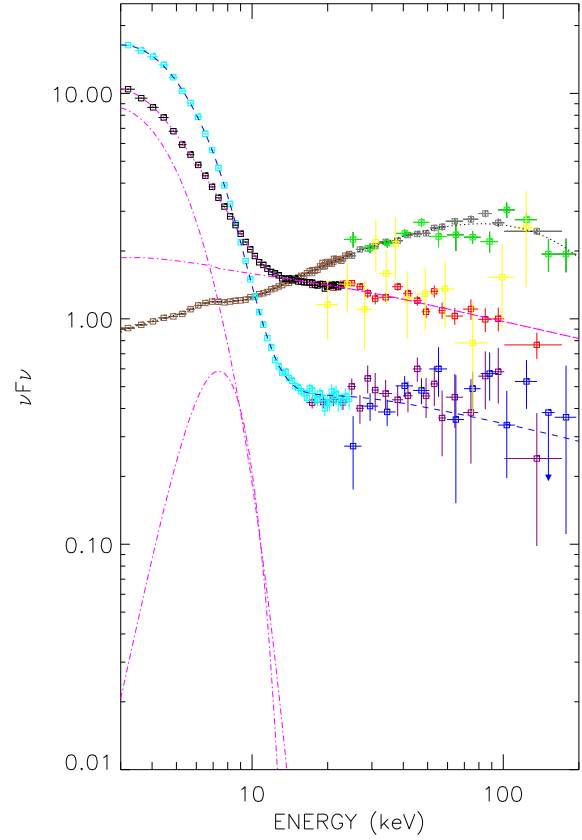}
\end{center}
\caption{Spectra of GRO~J1655$-$40 from
PCA, SPI and HEXTE data fitted simultaneously
with the model described in Table 4.
PCA data for revolutions 293 (brown), 294-B (black), 295+296 (indigo);
HEXTE data for revolutions 293 (gray), 294-B (red), 295+296 (violet);
SPI data for revolutions 293 (green), 295+296 (blue)
and BAT data for revolution 294$_{sw}$-B (yellow).}
\label{fig:fig5}
\end{figure}

From the dataset 294-C for which only PCA and HEXTE data are
available, there is no evidence of a high-energy cutoff up to 286
keV.
The reduced $\chi ^2$ decreases from 2.00(61 dof) down to
1.81(61 dof) and from 1.66(62 dof) down to 1.40(62 dof) for the datasets 294-C
and 295-A, respectively, using $N_H = 5 \times 10^{21} \rm \, cm^{-2}$. The high reduced $\chi ^2$ is due to poor modeling of
the low energy part of the spectra. A complex spectral feature
around 7 keV has been fitted by Diaz Trigo et al.
 (2007), using XMM-Newton and \textit{INTEGRAL} data.
The spectra corresponding to these fit parameters (Table~\ref{tab:tab3})
are shown in Figure~\ref{fig:fig4} and~\ref{fig:fig5}.


\subsubsection{Reflection model PEXRAV}

We fitted all data with a reflection model, PEXRAV in XSPEC
(Magdziarz \& Zdziarski 1995), consisting of a power-law with a high
energy cut-off and reflection from neutral medium (see Table~\ref{tab:tab5}).
During the LHS, as the luminosity in the 3-600 keV
energy range increased by a factor of $\simeq$ 2.8 from revolution 289 up to
revolution 292, the spectral index is constant ($\Gamma$ $\simeq$ 1.3-1.4).
Similar values of the spectral index and  of the high
 energy cutoff are found using the CUTOFFPL and the PEXRAV models:
it is explained by the fact that the reflection fraction is so low
that an additional reflection component is not required in the fit.
The range of values for the high energy cutoff (180-380 keV)
are consistent with the results of Shaposhnikov et al. (2007) ($196
\pm 48$ keV) for the revolution 289. We determine an upper limit on
the energy cutoff when the source is into the HSS : $E_c >2000$ keV
for the datasets (295+296)$_{all}$ (see Table 5).

During all observations (from the LHS to the HSS),
 the  reflection component
is not constrained and does not exceed $\Omega = 0.2$, which is
lower than the value of $\Omega \simeq 0.5$ found by Shaposhnikov et al. (2007).

\subsubsection{Comptonization model COMPTT}

The hard power law plus cut-off model of the LHS is usually
interpreted as thermal Comptonisation in a hot ($kT_e \sim 100 \rm \, keV$)
optically thin plasma (the corona). We used the COMPTT model (Titarchuk
1994) in order to describe the high
 energy  spectrum. The inner temperature of the disk (T$_{in}$) of
 the multicolor
 disk blackbody model is forced to be equal to the soft photon temperature
(T${_0}$) of the Comptonization model. We see in Table~\ref{tab:tab6} that the
 optical depth is $\tau$ $\simeq 1.8-2.0$ while the temperature decreases
from $kT_e$ $\simeq 40$ down to $kT_e$ $\simeq 30$ keV from the LHS to the HIMS.
By freezing the temperature to $kT_e=$ 37 keV which is the value obtained by
Shaposhnikov et al. (2007) for the dataset corresponding to the
\textit{INTEGRAL} revolution 289, we determined an optical depth  of
$\tau=1.98^{+0.05}_{-0.05}$ which is lower by a factor of 2.2 compared to
the value found in this reference. Moreover the inner disk T$_{in}$
temperature is higher  : T$_{in} =$ 0.83 keV instead of T$_{in} =$ 0.60 keV.
This could be explained by the fact that the continuum above 100 keV
constrained by the \textit{INTEGRAL} data is different in this
reference.

Regardless of the fit models, we can describe the spectral evolution
in terms of an evolution of the geometry based on the Compton
parameter $y = (4 kT / m_e c^2) {\rm max} (\tau, \tau^2)$ (where the electronic
temperature kT and the optical depth $\tau$ are derived from a
Comptonization model, Sunyaev \& Titarchuk, 1980). During
the LHS, the Compton parameter is not varying significantly ($y \simeq$ 0.6). It starts to decrease from the
HIMS (revolution 293), and is correlated with the increase of the
normalization of the disk component. It decreases (see value in
Table~\ref{tab:tab6}) by a factor of 4.3 between revolution 289 and
revolution 294-B when the source changes into the SIMS. The LHS to the
Intermediate state (both HIMS and SIMS) transition
 corresponds to a gradual decrease of the inner radius of the cold
  accretion disk, associated  with either the cold disk penetrating
  the hot inner flow, or the  latter collapsing into an optically thick
accretion disk with small active   regions of hot plasma on top of it
   (Zdziarski et al. 2002). The enhanced soft photon flux from the disk
   tends to cool down the hot phase, leading to softer spectra. This can be
directly observed from the evolution of the blackbody component with
respect to the powerlaw component
   (see the ratio $\phi _{p} / \phi _{b}$ given in Table~\ref{tab:tab3}).

\section{Discussion}

\subsection{High energy cutoff and jet quenching}

We determined that all datasets in the low hard state require a high
energy cut-off. In addition to F-tests that indicate that the cut-offs
are required, we have produced contour plots of cut-off versus
power-law index for each dataset and those also
substantiated the existence of  the cut-off in single revolutions. In
Figure 5, the contour plot corresponding to the revolution
290 clearly shows the requirement of the high energy cutoff during this
observation period which has also been studied by Caballero-Garcia et al. (2007)
who claimed that no cut-off is required for this dataset.
 The main difference between our analysis and theirs is the fact that
we have used the PCA to constrain the low energy part of the spectrum
whereas in Cabarello Garcia et al. (2007) JEM-X has been used.

For PEXRAV and CUTOFF models, the data from datasets 294-B and 294-C
indicate that either the high energy cutoff
 increases significantly, or the cutoff vanishes completely.
Such evolution was observed for GX339-4
in the HIMS (Belloni et al. 2005), and now appeared in the SIMS
in the case of GRO~J1655$-$40.

We attempt to describe this high energy feature by adding a powerlaw
component to the Comptonization model COMPTT. We found that the
addition of a power-law component leads to an improvement of the fit
significant at the $99.9998 \%$ level for both datasets according to
a F-test, with a best fit photon index of
 $\Gamma$=$3.86^{+0.12}_{-0.18}$ and $\Gamma$=$3.95^{+0.40}_{-0.35}$ for the data sets 294-B and 294-C, respectively.
 The powerlaw component contributes at $23\%$ and $20\%$ of
the 2-20 keV flux for the revolution 294-B and 294-C, respectively.
 The extension of a powerlaw at high energy has already been observed in
the Steep Power Law State of GRO~J1655$-$40 during its 1996 outburst
(see Remillard \& McClintock 2006). It is  interpreted as inverse
Compton scattering that occurs in a non-thermal corona, which may be
a simple slab operating on seed photons from the underlying disk (
Zdziarski et al. 2005). The origin of the Comptonizing electrons is
still a subject of debate. Poutanen \& Fabian (1999)
suggested that flare regions erupting from magnetic
instabilities in the accretion disk could  explain such a
Comptonizing media. The association of a non-thermal process with a
state transition has also been discussed for Cyg~X-1 (Malzac et al.
2006), in this case interpreting the power-law tail as the
Comptonization of soft photons by accelerated electrons (see also
Titarchuk 1997). Finally, such a high energy emission could also
come from an optically thick electron-positron outflow covering the
whole inner region of the accretion disk (Beloborodov 1999b).

\begin{figure}
\begin{center}
\includegraphics[scale=0.3,angle=270]{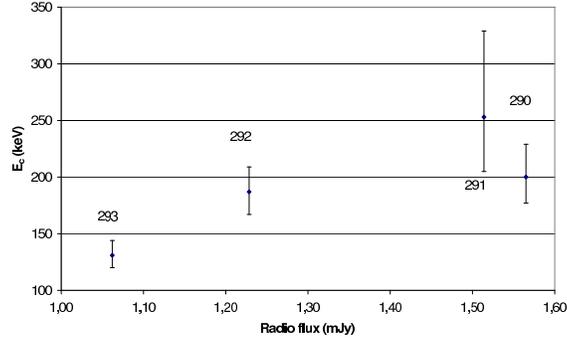}
\end{center}
\caption{Evolution of the high energy cutoff ($E_c$ expressed in keV)
derived from the model given in Table 4, as a function of the
radio flux measured by VLA at 8.460 GHz (Shaposhnikov et al. 2007). The
\textit{INTEGRAL} revolution numbers have been mentioned on the graph. There are no radio
data available after revolution 293.}
\label{fig:fig7}
\end{figure}

As seen from the Figure~\ref{fig:fig7}, there is a correlation
between the radio intensity and the exponential energy cutoff. The
maximum flux in radio is achieved when the source is in the LHS.
When the source enters the HIMS, the high energy cut-off
decreases (or the plasma cools down) as well as the radio flux.
Quenching of the jet in the soft states of black hole binaries is
known for many sources (Fender et al. 1999, Corbel et al. 2000).
Here we show a clear evolutionary path to this quenching in terms of
plasma temperature, perhaps associated with the cooling and
shrinking of the corona.

\subsection{The jet ejection event}

The jet ejections usually occur at or near state
transitions prior to the source entering the Intermediate State (Corbel et al.
2004). A radio flare ejection is observed for GRO~J1655$-$40 during
the revolution 291, just before the X- and $\gamma$-ray flux peak.
Such a behavior has also been observed for the black hole candidate
H1743-322 and GX339-4 (Joinet et al. 2005, Corbel et al. 2000,
Fender et al. 1999) and could be linked to a coronal ejection event.
The decrease of the Compton y-parameter (see Table 6), and the
appearance of the non thermal component ( addition of the powerlaw
component) indicate large changes in the corona geometry.

 Moreover, the small reflection fraction combined with
a high coronal temperature (or high energy cutoff) can be
explained by the model of
Beloborodov (1999a)
which argues that the corona above the accretion
disk is fed by magnetic flares or by a non-static coronae
(Malzac et al. 2001).  GRO~J1655$-$40 started to enter the IS from the end of
 the
revolution 292. This state corresponds to the decay of the radio emission
(Figure 4) which is usually dominated by the decaying optically thin synchrotron
emission from the jet ejections. During this state, such an emission
is decoupled from the black hole system as the emitting electrons are far from
it. The radio emission is completely quenched starting from the SIMS (revolution
294-B).

\section{Conclusion}

We clearly observed a transition from the LHS to the HSS during the
rising phase of the 2005 outburst of GRO~J1655$-$40 between MJD 53425 and
 MJD 53445. The outburst was covered with SPI, HEXTE, PCA and BAT
which allowed us to determine the high energy cutoff with a high precision.
During revolution 294, only data from the \textit{RXTE} and
\textit{Swift} observatory were available to constrain the high energy
cutoff during an intermediate state. An evolution of the high energy feature
 was noticed during the LHS : it decreased from a value of about 200 keV down to
130 keV when the source reached the maximum of luminosity in the $\gamma$-ray
domain (above 23-600 keV). For the HSS, a lower limit for the
high energy cutoff has been determined. We deduced that
there is no cut-off detectable in this spectral state.
This decrease
 corresponds to a decrease in the radio flux, and the cut-off disappeared
along with the radio jet. Finally, we also discussed on the relation
between state transition and the emission of jet in X-ray binaries
which is a subjet of significant importance in order to have a complete
view of the X-ray geometry. It would be interesting to perform a
broadband spectral fitting (from radio up to MeV) in order to
discriminate the different non thermal processes suggested in
this study.

\acknowledgments

The SPI project has been completed under the responsibility and leadership
of the CNES. We are grateful to ASI, CEA, DLR, ESA, INTA, NASA and OSTC
 for support. \\
Specific softwares used for this work have been developed by L.
Bouchet. A. J. thanks N. Shaposhnikov and M. Rupen for all
informations concerning the radio detection and M.D.
Caballero-Garcia who provided the ISGRI data used in this paper. E.
K. acknowledges support of TUB\.ITAK Career Program 106T570, Turkish
National Academy of Sciences Young and Successful Scientist Award
and Marie Curie International Reintegration Grant
MIRG-CT-2005-017203. Finally the authors are grateful to the anonymous referee for the very fruitful comments.

\clearpage
\thispagestyle{empty}
\setlength{\voffset}{-15mm}
\begin{table}[htp!]
\centering
\begin{tabular}{lcccccccc} 
\hline
Rev.  &SP$_{start}$ &SP$_{stop}$& $\Delta$t$_{sp}$ (ks)&ID&RX$_{start}$& RX$_{stop}$&  Exp.(ks)\\
\hline
\hline
289    &53425.14   &53427.36    &   133             & 90058-16-04-00 &  53425.06          &   53425.10         & 3.7          \\
       &           &            &                   & 90428-01-01-00 & 53426.04          &    53426.28         & 20.6          \\
       &           &            &                   & 90058-16-05-00 & 53427.02          &   53427.06         & 3.1          \\
290       & 53428.13           &  53430.36          &     134              & 90428-01-01-03 & 53428.14         &   53428.20         & 5.1          \\
         &                             &            &                   & 90428-01-01-04 & 53428.86         &    53429.12         & 22.5          \\
          &                     &                    &                     & 90428-01-01-02 &  53429.71         &   53429.97        & 22.6          \\
290$_{ibis}$ (A)& 53428.20           &  53429.50          &     69              &  &        &      &           \\
291    &53432.85   &53433.47    &   29             & 90428-01-01-10 & 53432.79          &   53433.00       & 17.97          \\
292 & 53434.80 & 53436.42   &     36           & 91404-01-01-02 & 53433.91          &  53434.09        & 16.14          \\
    &  &    &                & 91404-01-01-03 & 53434.69          &   53434.73        & 2.80          \\
    &  &    &               & 91404-01-01-01 &  53435.61         &   53435.64        &2.37          \\
    &  &    &           & 91404-01-01-04 & 53436.16          &   53436.17      &1.41          \\
292$_{sw}$-A,B& 53434.89   &53436.45&17.75 & & & & \\
\hline
293a      & --          &  --           &        --         & 91702-01-01-00 & 53436.72         &   53436.81         & 7.4         \\
293$_{sw}$      & 53436.48          &  53436.49          &   14.89            &   &          &            &           \\
293      & 53437.11          &  53438.34           &         51          & 91702-01-01-03 & 53438.05         &   53438.08         & 2.2          \\
294a        & -- & -- & -- & 91704-04-01-00&53439.61 & 53439.65 & 3.4 \\
294b         & -- & -- & -- & 91704-04-01-01& 53439.74 & 53439.78 & 3.6 \\
294$_{sw}$-A   & 53439.05 & 53439.65 & 13.29  & & & & \\ 
\hline
294$_{sw}$-B & 53440.72 & 53441.87 &4.42 & & & & \\
294-B       & --        & --         &    --             & 91702-01-02-01 & 53441.51            &     53441.54      &  2.3          \\
294-B       &  --         &   --         &   --                & 91702-01-02-02 &         53441.59 &      53441.60        &      1.4    \\
294-B       &  --         &   --         &    --               & 91702-01-02-03 &        53441.98    &      53442.01       &   2.2         \\
 294-C      &  --         &   --         &   --                & 91702-01-02-04 &       53442.06    &    53442.07          &    1.8       \\

294-C      &   --        &    --        &    --               & 91702-01-02-05 &       53442.12    &    53442.14         &    1.5      \\
294-C       &   --        &   --         &   --                & 91702-01-02-06 &       53442.58    &    53442.66          &    6.9       \\

\hline
295-A & -- & --&--& 91702-01-03-00&53443.54&53443.80&22.5\\
(295+296)$_{sw}$ &53444.92& 53447.21 & 15.33& & & & \\
295+296   &53445.07    &53447.71    &   75             & 91702-01-04-01 &  53444.49          &   53444.50         & 0.6          \\
(295+296)$_{ibis} (A)$ & 53445.10           &  53447.80          &     72             &  &        &      &           \\

\hline
\end{tabular}
\caption{The INTEGRAL observations of GRO J1655-40.
For each INTEGRAL revolution (Rev.), we give the beginning 
SP$_{start}$  and the end SP$_{stop}$ of the INTEGRAL observations 
in MJD from SPI detector.
Revolutions 295 and 296 have been merged (295+296).
(A) As the observation period  was not the same
for SPI and IBIS detectors, we indicate this first one for
IBIS/ISGRI data : revolutions 290$_{ibis}$ and 295+296$_{ibis}$.
$\Delta$t$_{sp}$ is the useful duration for 
INTEGRAL observations. 
ID is the identification program number of RXTE observations. RX$_{start}$ and RX$_{stop}$ are
the beginning and the end of RXTE observations taken quasi simultaneously with INTEGRAL observations.
Exp. is the exposure time for PCA. We indicate 
the simultaneous BAT data: 
the symbol "sw" is attached to the number of the revolution. 
The details of the BAT observations are given in Table 2. We separated 
with a line the observation period for which the source harboured 
the same X-ray state (see section 3.1 and Figure 1). This will be done for all Tables in this article.} 
\label{tab:tab1}
\end{table}

\clearpage
\setlength{\voffset}{0mm}
\begin{table}[htp!]
\centering
\begin{tabular}{lccccc} 
\hline
Rev.& ID         &  Sw$_{start}$ & Sw$_{stop}$  & Exp(ks)\\
\hline
\hline
292$_{sw}$-A &  00106709002  &53434.89  &53434.91 & 1.35\\
    &  00106709003  &53435.23  &53435.24  &1.35\\
    &  00030009002  &53435.42  &53435.50  &2.94\\
    &  00106709004  &53435.76 & 53435.78 & 0.90\\
    &  00055750001  &53436.08 & 53436.09  &0.90\\
    &  00055750002  &53436.15  &53436.16 & 1.30\\
    &  00055750003  &53436.21 & 53436.23  &1.35\\
292$_{sw}$-B    &  00106709005  &53436.23  &53436.25 & 1.35\\ 
    &  00055750004  &53436.28&  53436.30 & 1.35\\
    &  00058739002  &53436.30  &53436.32 & 1.66\\
    &  00055750005  &53436.35 & 53436.35&  0.64\\
    &  00058739002  &53436.37 & 53436.38  &1.04\\
    &  00055750006  &53436.41  &53436.42 & 0.64\\
    &  00058739002  &53436.44 & 53436.45 & 0.98\\
\hline
293$_{sw}$ & 00055750007 & 53436.48 & 53436.49 & 0.64 \\ 
	& 00055750008 & 53436.54 & 53436.55 & 0.45 \\ 
	& 00107547001 & 53436.62 & 53437.17 & 18.30 \\     
294$_{sw}$-A   & 00107547002 & 53439.05 & 53439.65 & 13.29 \\ 
\hline
294$_{sw}$-B      & 00058736001 & 53440.72 & 53440.93 & 1.80\\ 
	& 00058746001 & 53440.98 & 53441.00 & 1.80 \\ 
        & 00058746002 & 53441.07 & 53441.87 & 0.82 \\ 
\hline
(295+296)$_{sw}$& 00055800001 & 53444.92 & 53445.13 & 1.24 \\ 
 & 00111063001 & 53446.14 & 53446.88 & 12.60 \\ 
    & 00058752001 & 53447.06 & 53447.21 & 1.49\\ 
\hline
\end{tabular}
\caption{Details about BAT data observations.
We give the observation ID number,
the beginning Sw$_{start}$  and the end Sw$_{stop}$ of the BAT  
observations in MJD. Exp is the net exposure time. Revolutions 295 and 296
has been merged and is named (295+296)$_{sw}$.} 
\label{tab:tab2}
\end{table}

\clearpage
\begin{table}[ht]
\renewcommand{\arraystretch}{0.01}
\begin{tabular}{lccccc} 
\hline
rev   &  $\Phi$  [23-51 keV] &  $\Phi$    [51-95 keV]   &  $\Phi$ [95-160 keV] &$\Phi$ [160-270 keV]  \\	

\hline
\hline
289	&$42^{+2}_{-2}$ & $64^{+6}_{-6}$ &  $81^{+7}_{-7}$ & $59^{+16}_{-16}$ \\
290	&$49^{+4}_{-4}$ & $74^{+9}_{-9}$ &  $89^{+7}_{-7}$ &$87^{+21}_{-21}$ \\
291	&$50^{+14}_{-14}$ & $116^{+30}_{-30}$ &  $88^{+30}_{-30}$&$167^{+70}_{-70}$ \\
292	&$125^{+8}_{-8}$ & $159^{+19}_{-19}$ &  $245^{+25}_{-25}$&$187^{+50}_{-50}$ \\
\hline
293$_{sw}$ 	&$241^{+3}_{-4}$ & $320^{+7}_{-6}$ &  $403^{+12}_{-12}$&-- \\
293	&$280^{+7}_{-7}$ & $327^{+13}_{-13}$ &  $358^{+14}_{-14}$&$214^{+37}_{-37}$ \\
294$_{sw}$-A	&$328^{+5}_{-5}$ & $357^{+7}_{-7}$ &  $382^{+14}_{-12}$&-- \\
\hline
294$_{sw}$-B 	&$367^{+6}_{-7}$ & $347^{+11}_{-10}$ &  $332^{+17}_{-17}$&-- \\
\hline
(295+296)$_{sw}$	(A) &$81^{+3}_{-3}$ &$76^{+6}_{-6}$ &  $73^{+10}_{-9}$&- \\
\hline
\end{tabular}
\caption{Flux $\Phi$ (expressed in mCrab) of GRO J1655-40 
measured by SPI during different INTEGRAL revolutions (rev) and
for several energy bands. (A) These data correspond to 
revolutions 295 and 296 which have been merged.}
\vspace*{-0.3 cm} 
\label{tab:tab4}
\end{table}

\clearpage
\begin{table}[ht]
\renewcommand{\arraystretch}{0.01}
\begin{tabular}{lcccccccc}
\hline
rev   &  T$_{in}$ & N$_{in}$ &$\Gamma$       &  E$_{c}$     & W$_{Fe}$ &$\chi ^2$(dof)& F-test&$\phi _p / \phi _b$\\
      &    keV&       &              &   keV        &  eV      &              &  &\\
\hline
\hline
289 & $1.29^{+0.12}_{-0.12}$   & $0.80^{+0.37}_{-0.23}$& $1.47^{+0.02}_{-0.02}$ &--- &$76^{+67}_{-60}$ &1.86 (79)& &\\
289 & $1.41^{+0.08}_{-0.14}$   & $1.41^{+0.49}_{-0.17}$& $1.36^{+0.04}_{-0.04}$ &$231^{+94}_{-50}$&$75^{+42}_{-48}$ &0.95(78) & 3.62E-10&9.6\\
290 & $1.20^{+0.10}_{-0.05}$   & $1.05^{+0.58}_{-0.31}$& $1.47^{+0.01}_{-0.01}$ &--- &$75^{+36}_{-50}$ &2.04(79)&&\\
290 & $1.40^{+0.08}_{-0.09}$   & $1.21^{+0.33}_{-0.22}$& $1.33^{+0.03}_{-0.03}$ &$259^{+62}_{-41}$ &$76^{+42}_{-40}$ &0.86(78)&1.85E-16&9.6\\
290 (A) & $1.09^{+0.12}_{-0.11}$   & $1.15^{+0.73}_{-0.44}$& $1.51^{+0.01}_{-0.01}$ &--- &$69^{+41}_{-41}$ &3.45(108)&&\\
290 (A) &  $1.39^{+0.07}_{-0.08}$  & $1.32^{+0.34}_{-0.23}$& $1.32^{+0.01}_{-0.03}$ &$200^{+29}_{-23}$ &$77^{+58}_{-77}$ &1.37(107)&1.69E-23&9.7\\
291 & $1.04^{+0.14}_{-0.15}$   & $2.64^{+3.27}_{-1.15}$& $1.46^{+0.01}_{-0.01}$ &--- &$75^{+36}_{-50}$ &1.60(79)&\\
291 & $1.40^{+0.12}_{-0.16}$   & $1.37^{+0.82}_{-0.32}$& $1.34^{+0.03}_{-0.04}$ &$253^{+76}_{-48}$ &$78^{+42}_{-42}$ &0.85(78)&1.42E-12&8.7\\
292 & $1.09^{+0.08}_{-0.08}$   & $4.32^{+1.87}_{-1.24}$& $1.48^{+0.01}_{-0.01}$ &--- &$113^{+72}_{-82}$ &4.48(77)&&\\

292 & $1.27^{+0.09}_{-0.08}$   & $4.49^{+1.42}_{-1.12}$& $1.29^{+0.02}_{-0.02}$ &$187^{+22}_{-20}$ &$177^{+88}_{-81}$ &0.85(76)&2.27E-29&7.5\\
292 + $292_{sw}$    & $1.27^{+0.09}_{-0.08}$   & $4.50^{+1.42}_{-1.12}$& $1.29^{+0.02}_{-0.02}$ &$186^{+22}_{-18}$ &$76^{+42}_{-40}$ &0.68(223)&&\\
\hline
293$_{sw}$ + 293a   & $0.93^{+0.06}_{-0.06}$  & $22^{+9}_{-6}$ & $1.53^{+0.01}_{-0.01}$ &--&$145^{+75}_{-52}$ &3.85(118)&--&--\\
293$_{sw}$ + 293a   & $1.07^{+0.05}_{-0.05}$  & $22^{+5}_{-4}$ & $1.33^{+0.02}_{-0.02}$ &$173^{+21}_{-17}$ &$315^{+93}_{-57}$ &1.35(117)&1.19E-28&7.4\\
293 & $1.60^{+0.05}_{-0.01}$   & $4.87^{+1.52}_{-0.76}$& $1.62^{+0.05}_{-0.05}$ &--- &$388^{+46}_{-39}$ &5.30(77)&&\\
293 & $1.00^{+0.03}_{-0.03}$   & $45^{+21}_{-15}$& $1.37^{+0.02}_{-0.03}$ &$131^{+13}_{-11}$ &$386^{+80}_{-52}$ &1.15(76)&3.46E-27&5.3\\
294$_{sw}$-A + 294a,b   & $0.67^{+0.03}_{-0.03}$   & $286^{+17}_{-14}$ & $1.82^{+0.3}_{-0.02}$ &-- &$166^{+28}_{-38}$ &14.32(118)&--& \\
294$_{sw}$-A + 294a,b   & $1.01^{+0.03}_{-0.03}$   & $121^{+17}_{-14}$ & $1.41^{+0.3}_{-0.02}$ &$87^{+4}_{-5}$ &$646^{+67}_{-46}$ &1.37(117)&--&4.4\\
\hline
294-B+294$_{sw}$-B  & $0.98^{+0.01}_{-0.01}$   & $1681^{+64}_{-66}$& $2.18^{+0.01}_{-0.02}$ &-- &$622^{+35}_{-44}$ &1.16(118)& &\\
294-B+294$_{sw}$-B  & $0.98^{+0.01}_{-0.01}$   & $1687^{+65}_{-61}$& $2.08^{+0.05}_{-0.04}$ &$302^{+225}_{-81}$ &$673^{+50}_{-41}$ &0.99(117)&1.682E-05 &0.41\\
294-C   & $1.02^{+0.01}_{-0.01}$   & $1686^{+54}_{-67}$& $2.15^{+0.01}_{-0.01}$ &-- &$522^{+174}_{-108}$ &1.96(62)& &\\
294-C   & $1.02^{+0.01}_{-0.01}$   & $1676^{+61}_{-57}$& $2.08^{+0.02}_{-0.04}$ &$439^{+418}_{-153}$ &$538^{+35}_{-49}$ &1.81(61)& 1.49E-02&0.29\\
\hline
295-A    & $1.09^{+0.01}_{-0.01}$   & $1709^{+60}_{-47}$& $2.02^{+0.01}_{-0.01}$ &$>800$  &$340^{+38}_{-37}$ &1.40(62)&&0.15\\
295-B+295$_{sw}$     & $1.15^{+0.01}_{-0.01}$   & $1454^{+63}_{-29}$& $1.85^{+0.07}_{-0.05}$ &$>261$&$103^{+50}_{-50}$ &1.03(135)&&0.05\\
295-B+295$_{sw}$     & $1.15^{+0.01}_{-0.01}$   & $1464^{+37}_{-27}$& $1.87^{+0.04}_{-0.05}$ &--- &$110^{+91}_{-80}$ &1.02(136)&&0.05\\
(295+296)$_{all}$ (A)& $1.13^{+0.01}_{-0.01}$   & $1760^{+20}_{-9}$& $2.17^{+0.01}_{-0.01}$ &---&$188^{+52}_{-43}$ &0.86(167)&&0.07\\
(295+296)$_{all}$ (A)& $1.14^{+0.01}_{-0.01}$   & $1649^{+231}_{-134}$& $2.05^{+0.11}_{-0.07}$ &$>237$ &$167^{+46}_{-44}$ &0.84(166)&&0.05\\
\hline
\end{tabular}
\caption{PCA, SPI and HEXTE data fitted simultaneously
using the XSPEC multicomponent model
PHABS*(GAUSSIAN+DISKBB+POWERLAW).
T$_{in}$ is
 the inner disk temperature and N$_{in}$ the normalisation,
 a Gaussian line was fixed at an energy of 6.4 keV
with a width fixed to 0.1 keV for rev 289, 290.
W$_{Fe}$ is the equivalent width.
The interstellar absorption PHABS has been fixed to $0.7 \times 10^{22}$ cm$^{-2}$
until the dataset 294$_{sw}$-A + 294a,b and to $0.5 \times 10^{22}$ cm$^{-2}$ from the revolution
294-B onward.
$\Gamma$ is the photon index.
The POWERLAW component was replaced by a CUTOFFPL component and
 E$_c$ is the high energy cut-off.
 The F-test is calculated between the POWERLAW and CUTOFFPL models.
We also give the ratio $\phi _p / \phi _b$ where $\phi _p $ and
$\phi _b$ are the powerlaw and the blackbody  flux in the 2-20 keV
energy range respectively. $\chi ^2$(dof) is the reduced $\chi ^2$
with the degree of freedom (dof). For revolutions 295 and 296 which
have been merged, the SPI (295+296) and Swift data (295+296)$_{sw}$
have been used. This data set is named (295+296)$_{all}$. (A) For
revolution 290 as well as for revolutions (295+296)$_{all}$, the
IBIS/ISGRI data have been added (290$_{ibis}$ and
(295+296)$_{ibis}$).} \vspace*{-0.3 cm} \label{tab:tab3}
\end{table}

\clearpage
\begin{table}[ht]
\renewcommand{\arraystretch}{0.01}
\begin{tabular}{lcccccccc}
\hline
rev   &   $\Gamma$       &  E$_{c}$  &  T$_{in}$ &N$_{in}$  & W$_{Fe}$              &$\chi ^2$(dof)&L$_{3-600}$ $\times$ 10$^{-9}$  \\
      &                  &  keV      &   keV     &          &                  eV    &            &             ergs cm$^{-2}$ s$^{-1}$\\
\hline
\hline
289    &$1.36^{+0.04}_{-0.08}$ &$237^{+143}_{-55}$&$1.38^{+0.12}_{-0.11}$&$1.07^{+0.48}_{-0.27}$  &$80^{+65}_{-78}$&0.98(77)&$2.9^{+0.1}_{-0.1}$\\
290    &$1.33^{+0.01}_{-0.06}$ &$255^{+108}_{-35}$&$1.44^{+0.05}_{-0.11}$&$1.07^{+0.26}_{-0.15}$                     &$71^{+55}_{-61}$&0.89(77)&$3.5^{+0.1}_{-0.1}$\\
290 (A)   &$1.33^{+0.08}_{-0.01}$ &$207^{+111}_{-16}$&$1.32^{+0.04}_{-0.01}$&$1.68^{+0.56}_{-0.51}$                     &$86^{+62}_{-78}$&1.40(106)&$3.3^{+0.02}_{-0.2}$\\
291    &$1.37^{+0.10}_{-0.02}$ &$316^{+232}_{-23}$&$1.36^{+0.17}_{-0.26}$&$2.36^{+3.81}_{-0.90}$                     &$<130$&0.85(77)&$6.4^{+0.5}_{-0.6}$\\
292    &$1.29^{+0.03}_{-0.02}$ &$185^{+30}_{-19}$&$1.27^{+0.10}_{-0.08}$&$5.73^{+1.80}_{-1.43}$                   &$<173$&0.93(77)&$8.1^{+0.4}_{-0.1}$\\
\hline
293$_{sw}$ + 293a    &$1.33^{+0.07}_{-0.02}$&$171^{+59}_{-15}$&$1.07^{+0.05}_{-0.05}$&$22^{+5}_{-4} $                  &$317^{+80}_{-50}$&1.36(116)&$10.6^{+0.6}_{-0.3}$\\
293    &$1.37^{+0.05}_{-0.05}$ &$131^{+14}_{-8}$&$1.04^{+0.06}_{-0.04}$&$48^{+18}_{-16} $                  &$381^{+88}_{-82}$&1.22(77)&$13.9^{+0.5}_{-0.1}$\\
294$_{sw}$-A + 294a,b    &$1.42^{+0.05}_{-0.05}$&$88^{+14}_{-8}$&$1.00^{+0.06}_{-0.04}$  &$121^{+18}_{-16} $                  &$645^{+59}_{-56}$&1.39(16)&$15.35^{+0.1}_{-0.1}$\\
\hline
294-B+294$_{sw}$-B     &$2.07^{+0.12}_{-0.01}$ &$289^{+151}_{-36}$&$0.97^{+0.01}_{-0.01}$  &$1752^{+24}_{-86} $                  &$701^{+100}_{-75}$&1.06(116)&$18.13^{+0.03}_{-0.03}$\\
\hline
295-A   &$2.03^{+0.05}_{-0.01}$ &$>700$&$1.09^{+0.01}_{-0.01}$  & $1709^{+55}_{-56}$&$>156$&1.45(60)&$23.39^{+0.25}_{-0.16}$\\
295-B+295$_{sw}$-B &$1.86^{+0.04}_{-0.12}$ &$>457$&$1.15^{+0.01}_{-0.01}$  & $1457^{+64}_{-32}$&$110^{+60}_{-61}$&1.03(134)&$22.19^{+0.32}_{-0.11}$\\
(295+296)$_{all}$ (A) &$2.16^{+0.04}_{-0.11}$ & $>2000$&$1.14^{+0.01}_{-0.01}$  & $1509^{+67}_{-58}$&$<840$&0.82(164)&$20.84^{+3.24}_{-1.72}$\\
\hline
\end{tabular}
\caption{PCA, HEXTE and SPI  data fitted simultaneously
using the XSPEC multicomponent model PHABS*(PEXRAV+GAUSSIAN+DISKBB).
$\Gamma$ is the photon index and E$_c$ the energy cut-off.
T$_{in}$ is
 the inner disk temperature and N$_{in}$ the normalisation.
 The interstellar absorption PHABS has been fixed to
  $0.7 \times 10^{22}$ cm$^{-2}$
until the dataset 294$_{sw}$-A + 294a,b and to $0.5 \times 10^{22}$ cm$^{-2}$
from the revolution
294-B onward.
The Gaussian line was fixed at an energy of 6.4 keV. W$_{Fe}$ is the equivalent width.
 the reflection fraction $\Omega$/${2\pi}$ has been found with an
 upper value ranging 0.1-0.2.
L$_{3-600}$  is the luminosity of the
 source in the 3-600 keV energy range.
 $\chi ^2$(dof) is the reduced $\chi ^2$ with the degree of freedom (dof).
For revolutions 295 and 296 which have been merged, the SPI
(295+296) and Swift data (295+296)$_{sw}$ have been used. This data
set is named (295+296)$_{all}$. (A) For revolution 290 as well as
for revolutions (295+296)$_{all}$, the IBIS/ISGRI data have been
added (290$_{ibis}$ and (295+296)$_{ibis}$).}
\vspace*{-0.3 cm}
\label{tab:tab5}
\end{table}

\clearpage
\begin{table}[ht]
\renewcommand{\arraystretch}{0.01}
\begin{tabular}{lcccccc}
\hline
rev   &   $kT $       &  $\tau$  &  T$_{in}$ &N$_{in}$ &  $\chi ^2$(dof)&y\\
      &      keV      &           &   keV&         &              &\\
\hline
\hline
289 &$43^{+16}_{-8}$ &$1.78^{+0.26}_{-0.37}$& $0.83^{+0.04}_{-0.04}$ &$12^{+3}_{-2}$ & 1.04(79)&$1.07^{+0.40}_{-0.23}$\\
290 &$40^{+6}_{-4}$ &$1.92^{+0.13}_{-0.14}$&$0.80^{+0.03}_{-0.03}$&$16^{+3}_{-2}$& 0.93(78)&$1.15^{+0.17}_{-0.12}$\\
290 (B) &$36^{+4}_{-3}$ &$1.98^{+0.14}_{-0.12}$&$0.76^{+0.01}_{-0.03}$&$18^{+}_{-}$& 1.82(106)&$1.12^{+0.12}_{-0.10}$\\
291 &$36^{+12}_{-6}$ &$2.07^{+0.15}_{-0.17}$&$0.78^{+0.07}_{-0.06}$&$34^{+10}_{-21}$&0.84(77)&$1.21^{+0.40}_{-0.21}$\\
292 &$35^{+2}_{-2}$ &$2.09^{+0.10}_{-0.07}$&$0.80^{+0.02}_{-0.04}$&$44^{+10}_{-4}$&1.36(77)&$1.20^{+0.07}_{-0.07}$\\
\hline
293$_{sw}$ + 293a   &$31^{+2}_{-1}$ &$2.15^{+0.1}_{-0.1}$&$0.77^{+0.02}_{-0.02}$&$106^{+16}_{-13}$&1.63(117)&$1.12^{+0.07}_{-0.08}$\\
293 &$31^{+2}_{-2}$ &$2.01^{+0.09}_{-0.09}$&$0.74^{+0.10}_{-0.12}$&$244^{+338}_{-93}$&1.31(77)&$0.98^{+0.06}_{-0.06}$\\
294$_{sw}$-A + 294a,b   &$26^{+2}_{-2}$ &$1.93^{+0.09}_{-0.09}$&$0.76^{+0.10}_{-0.12}$&$457^{+338}_{-93}$&2.06(117)&$0.76^{+0.06}_{-0.06}$\\
\hline
294-B+294$_{sw}$-B  &$282^{+38}_{-41}$ &$0.03^{+0.01}_{-0.01}$&$0.95^{+0.01}_{-0.01}$&$2065^{+51}_{-115}$&1.07(117)&$0.06^{+0.01}_{-0.01}$\\
294-B+294$_{sw}$-B (A)  &$59^{+33}_{-23}$ &$0.55^{+0.38}_{-0.21}$&$0.95^{+0.03}_{-0.02}$&$1183^{+200}_{-221}$&0.75(115)&$0.25^{+0.14}_{-0.14}$\\
294-C   &$320^{+54}_{-92}$ &$0.02^{+0.02}_{-0.01}$&$1.00^{+0.01}_{-0.01}$&$1967^{+75}_{-58}$&1.91(61)&$0.05^{+0.01}_{-0.03}$\\
294-C (A)   &$86^{+61}_{-38}$ &$0.43^{+0.51}_{-0.23}$&$1.06^{+0.03}_{-0.03}$&$1192^{+197}_{-172}$&1.26(59)&$<0.20$\\
\hline
295-B+295$_{sw}$-B  &$763^{+161}_{-610}$ &$<0.07$&$1.24^{+0.01}_{-0.01}$&$1267^{+52}_{-37}$&1.02(135)&$<0.06$\\
(295+296)$_{all}$ (B) &$445^{+348}_{-88}$ &$<0.12$&$1.17^{+0.01}_{-0.01}$&$1402^{+63}_{-48}$&1.04(165)&$<0.07$\\
\hline
\end{tabular}
\caption{PCA, HEXTE and SPI  data fitted simultaneously
using the XSPEC multicomponent model PHABS*(COMPTT+GAUSSIAN+DISKBB).
The interstellar absorption PHABS has been fixed to $0.7 \times 10^{22}$ cm$^{-2}$
until the dataset 294-A and to $0.5 \times 10^{22}$ cm$^{-2}$ from the revolution
294-B onward.
T$_{in}$
  is the inner disk temperature, N$_{in}$ the normalisation,
$\tau$ the optical depth and kT the plasma temperature.
 The Compton parameter y (see the definition in the text)
  has been determined. (a) For the dataset 294-B and 294-C
  a powerlaw component $\Gamma$ of $3.86^{+0.12}_{-0.18}$  and $3.95^{+0.40}_{-0.35}$
   was added, respectivelly.
  $\chi ^2$(dof) is the reduced $\chi ^2$ with the degree of freedom (dof).
  For revolutions 295 and 296 which have been merged, the SPI (295+296) and Swift data (295+296)$_{sw}$ have been used.
This data set was named (295+296)$_{all}$.
  (B) For revolution 290 as well as for revolutions (295+296)$_{all}$, the ISGRI data have been added (290$_{ibis}$ and (295+296)$_{ibis}$).}
\vspace*{-0.3 cm}
\label{tab:tab6}
\end{table}

\clearpage

\begin{figure}
\begin{center}
\includegraphics[scale=0.3,angle=270]{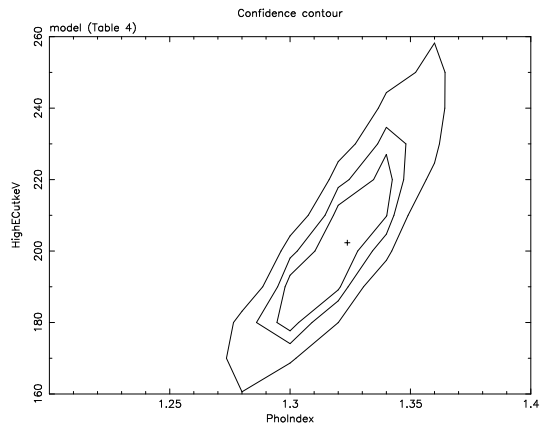}
\end{center}
\caption{$\Gamma$-$E_c$ contour plot (revolution 290) for the simultaneous PCA, HEXTE, ISGRI and SPI
 observations using the model described in Table 4 . The curves refer to
$\delta \chi ^2$ = 2.30, 4.61, 9.21, corresponding to confidence
levels of
68, 90 and 99\% for two interesting parameters.}
\end{figure}

\end{document}